\documentclass[prc,amsmath,superscriptaddress,showpacs,amssymb,floatfix,aps,twocolumn,a4paper]{revtex4}

\usepackage{amssymb}
\usepackage{epsfig}
\usepackage{ulem}
\usepackage[usenames]{color}
\usepackage{amsmath}
\usepackage{amsfonts}
\usepackage{lmodern,dsfont}

\newcommand{\be}{\begin{eqnarray}}
\newcommand{\ee}{\end{eqnarray}}
\newcommand{\non}{\nonumber \\}

\begin{document}

\title{Relevance of complex branch points for partial wave analysis}

\author{S. Ceci}
\affiliation{Rudjer Bo\v{s}kovi\'{c} Institute, Bijeni\v{c}ka  54, HR-10000 Zagreb, Croatia}
\author{M. D\"oring}
\affiliation{
HISKP (Theorie),Universit\"at Bonn, Nu\ss allee 14-16, D-53115 Bonn, Germany}
\author{C.~Hanhart} 
\affiliation{Institut f\"ur Kernphysik and J\"ulich Center for Hadron Physics, 
Forschungszentrum J\"ulich, D-52425 J\"ulich, Germany}
\affiliation{Institute for Advanced Simulation,
Forschungszentrum J\"ulich, D-52425 J\"ulich,Germany}
\author{S. Krewald}
\affiliation{Institut f\"ur Kernphysik and J\"ulich Center for Hadron Physics, 
Forschungszentrum J\"ulich, D-52425 J\"ulich, Germany}
\affiliation{Institute for Advanced Simulation,
Forschungszentrum J\"ulich, D-52425 J\"ulich,Germany}
\author{U.-G.~Mei{\ss}ner}
\affiliation{
HISKP (Theorie),Universit\"at Bonn, Nu\ss allee 14-16, D-53115 Bonn, Germany}
\affiliation{Institut f\"ur Kernphysik and J\"ulich Center for Hadron Physics, 
Forschungszentrum J\"ulich, D-52425 J\"ulich, Germany}
\affiliation{Institute for Advanced Simulation,
Forschungszentrum J\"ulich, D-52425 J\"ulich,Germany}
\author{A.~\v Svarc}
\affiliation{Rudjer Bo\v{s}kovi\'{c} Institute, Bijeni\v{c}ka  54, HR-10000 Zagreb, Croatia}


\begin{abstract}
A central issue in hadron spectroscopy is to deduce --- and interpret --- resonance parameters, namely pole positions and residues,
from experimental data, for those are the quantities to be compared to lattice QCD or model calculations. However, not every
structure in the observables derives from a  resonance pole: the origin might as well be branch points, either located on the real axis
(when a new channel comprised of stable particles opens) or in  the complex plane (when at least one of the intermediate particles is
unstable). In this paper we demonstrate first the existence of such branch points in the complex plane and then show on the example of
the $\pi N$ $P_{11}$ partial wave that it is not possible to distinguish the structures induced by the latter from a true pole signal
based on elastic data alone.
\end{abstract}


\maketitle


\section{Introduction}

The second and third resonance region of baryonic excited states is currently under intense experimental investigation at various
laboratories such as ELSA, MAMI, or JLab~\cite{Klempt:2009pi,Sparks:2010vb,Ahrens:2006gp,Aznauryan:2011ub}. Many resonances overlap at
these energies, and usually partial wave analyses in different frameworks, such as $K$-matrix approaches or dynamical coupled channel
models~\cite{Arndt:2006bf, Arndt:2008zz,Batinic:1995kr, Batinic:1997gk,Drechsel:2007if, Workman:2011hi,Penner:2002ma,Shklyar:2005xg,
Anisovich:2010an, Krehl:1999km, Gasparyan:2003fp,Doring:2009bi, Doring:2009yv,Doring:2010ap, Paris:2008ig, JuliaDiaz:2007kz,
Suzuki:2009nj,Tiator:2010rp} are necessary to disentangle the resonance content. Furthermore, many resonances may couple only weakly to
the $\pi N$ channel, and the investigation of different initial and final states in hadronic reactions is
mandatory~\cite{Doring:2010ap}. Also, at these energies multi-pion intermediate and final states are becoming increasingly important and
should be included in the analysis of the $S$-matrix. For the corresponding $T$-matrix, channels with stable particles like $\eta N$
induce a branch point at the threshold energy ($\sqrt{s}=m_\eta+M_N$), that may be visible as cusps in the
amplitude~\cite{Doring:2009uc,Doring:2009qr}. 

For effective multi-pion channels with one unstable and one stable particle, such as $\rho N$, the analytic structure is more
complicated.  In comparison to the branch points on the real $s$ axis and the first and second  sheet poles,  the third type of allowed
singularities are branch points within the complex energy plane. They emerge when amongst groups of particles of an at least three--body
decay there exists a strong correlations between two particles. For example, a significant fraction of $\pi^+\pi^- X$ intermediate and
final states typically goes through the $\rho$ meson. The resulting line shapes are discussed in Ref.~\cite{Hanhart:2010wh}. Branch 
points in the complex plane also emerge in the recently developed complex-mass scheme for baryonic resonances~\cite{Djukanovic:2010zz}.

Known theoretically for a long time~\cite{brapothree,Cutkosky:1990zh}, these branch points are present in several modern approaches, 
such as the GWU/SAID analysis~\cite{Arndt:2006bf,Arndt:2008zz}, the
J\"ulich~\cite{Krehl:1999km,Gasparyan:2003fp,Doring:2009bi,Doring:2009yv,Doring:2010ap} and EBAC~\cite{JuliaDiaz:2007kz,Suzuki:2009nj}
approaches, or the Bonn-Gatchina~\cite{Anisovich:2010an} analysis. It is the goal of this study to demonstrate the model-independent
character of those complex branch points. To do so, we employ general properties of the $S$-matrix only. In a particular example it is
then shown that the branch points are of relevance in partial wave analyses: if the theoretical partial wave does not include them,
their absence can easily be simulated by resonance poles. This, of course, distorts the extracted baryon spectrum. Branch points in the
complex plane are thus important for the reliable extraction of resonance parameters.

The paper is organized as follows: in Sec.~\ref{sec:anal} the existence of branch points in the complex plane is derived from 
three-body phase space, in Sec.~\ref{sec:threshold} the properties of the branch points are determined, and in Sec.~\ref{sec:relevance}
it is shown that these branch points are relevant in the extraction of the resonance content of partial waves.


\section{Analytic structure of the $S$-matrix and  complex branch points}
\label{sec:anal}
Every channel opening introduces a new branch point and with it a  new sheet to the $S$-matrix, located at $s=(\sum m_i)^2$, with $m_i$
the masses of the stable particles in that channel. The first sheet is always the physical one, i.e. where the physical
amplitude is situated. The only singularities allowed on the first sheet are poles on the real $s$ axis below the lowest threshold
(=bound states) or branch points on the real axis. On other sheets, poles and branch points can be located anywhere. Poles on the second
sheet are called resonances if their real part is located above the lowest threshold, and they are called virtual states, if they are
located below the threshold, but on the real axis. It is also possible to have poles on the second sheet inside the complex plane with a
real part lower than  the threshold~\cite{sigmamassdep}, or on other hidden sheets which are often referred to as shadow poles.

In this study we are interested in branch points on the second sheet in the complex plane, i.e. on the same sheet on which the resonance
poles are situated. To prove the emergence of these branch points, let us start from the optical theorem
\begin{multline}
T(j\to i)-T^\dagger(j\to i) \\
=i\,(2\pi)^4\sum_f
\int
d\Phi_f
T^\dagger(i\to f)
 T(j\to f) 
\label{opttheorem}
\end{multline}
where $T(j\to i)$ denotes the $T$-matrix connecting channels $i$ and $j$ and $d\Phi_f$ denotes the phase space of channel $f$. To
simplify the argument we assume that the $T$-matrix is in a particular partial wave; below we focus on the singularities that stem from
the unitarity cuts only. Singularities like the left-hand cuts, the short nucleon cut~\cite{hoehlerpin,Doring:2009yv} or the circular
cut, induced by the partial wave projection, are ignored in the following for they are irrelevant for the argument given. 

To be specific we use the normalization of phase space as proposed by the particle data group~\cite{Nakamura:2010zzi}.  Then we have for
the $n$--particle phase space
\begin{eqnarray}
d\Phi_n(P;p_1,..,p_n) = \delta^{(4)} \left(P-\sum_i p_i\right)\prod_i
\frac{d^3p_i}{(2\pi)^32E_i} 
\end{eqnarray}
where $P$ is the overall center-of-mass (c.m.) four-momentum.

To avoid complications, that are irrelevant for the validity of the present argumentation, we now focus on the diagonal channel $i=j$.
To be concrete we assume $i=\pi N$. To further simplify the argument, in addition we focus on $f=\rho N$ as the only relevant
intermediate $\pi\pi N$ channel. The latter assumption allows us to write
\begin{multline}
T(\pi N\to \pi\pi N)= iW(m_{\pi\pi}^2)D(m_{\pi\pi}^2)T(\pi N\to \rho N) ,
\end{multline}
where $D(m_{\pi\pi}^2)$ denotes the physical $\rho$ propagator as a function of the $\pi\pi$ invariant mass $m_{\pi\pi}$ and
$W(m^2_{\pi\pi})$ is the partial wave projected decay vertex, that contains also the information on the orbital angular momentum $\ell$
of the decay into $\pi\pi$. In the following we abbreviate $m\equiv m_{\pi\pi}$. 

One can decompose the three-body phase space into two subspaces~\cite{Nakamura:2010zzi},
\begin{multline}
d\Phi_n(P;p_1,..,p_n)= d\Phi_j(q;p_1,..,p_j)\\
\times\,
d\Phi_{n-j+1}(P;q,p_{j+1},..,p_n)(2\pi)^3dm^2 \ .
\label{decomp_phasespace}
\end{multline}
For the example of the $\rho[\pi\pi]\, N$ system considered here, the first factor $d\Phi$ refers to the $\pi\pi$ phase space in the
$\rho$ subsystem at four-momentum $q$ (note that $m^2=q^2$), the second is the $\rho N$ phase space at four-momentum $P$, and $n=3$,
$j=2$. With this decomposition,
\begin{multline}
\int d\Phi_j(q;p_1,..,p_j)|D(m^2)W(m^2)|^2 \\ = -\frac1{\pi}{\rm Im}(D(m^2))=\rho(m^2) \ ,
\end{multline}
where $\rho(m^2)$ denotes the spectral density for the resonance normalized via
\begin{equation}
\int_{4m_\pi^2}^\infty dm^2 \rho(m^2) = 1 \ .
\label{norm}
\end{equation}
We get for the discontinuity of the $\pi N$ amplitude from the $\pi\pi N$ channel
\begin{multline}
\frac1{i}\left(T(\pi N\to \pi N)-T^\dagger(\pi N\to \pi N)\right) \\ =
(2\pi)^7\int  dm^2 \rho(m^2)\int d\Phi_2(P;q,p_3) \\
\times\,|T(\pi N\to \rho N)(s,m^2)|^2 
 + ...  \ ,
\label{imT}
\end{multline}
where the ellipses denote contributions from the other channels omitted here. The two-body phase space can be calculated explicitly. One
finds
\begin{equation}
d\Phi_2(P;q,p_3) = \frac1{256\pi^6}\frac{p(\sqrt{s},m,m_3)}{\sqrt{s}}\ d\Omega \ ,
\end{equation}
with
\begin{multline}
p(\sqrt{s},m,m_3)\\
=\frac{1}{2\sqrt{s}}\sqrt{(s-(m_3+m)^2)(s-(m_3-m)^2)}
\label{qcm}
\end{multline}
for the c.m. momentum of the nucleon (particle 3) and the pion pair with invariant mass $m$.

Using Eq.~(\ref{imT}) we may thus express the $T$-matrix through a dispersion integral and obtain 
\begin{widetext}
\begin{equation}
T(\pi N\to \pi N) = \frac1{4} \int\limits_{(M_N+2m_\pi)^2}^\infty
\frac{ds'}{\sqrt{s'}}\int\limits_{4m_\pi^2}^{(\sqrt{s'}-M_N)^2}dm^2 
\,\rho(m^2)\,p(\sqrt{s'},m,m_3)\int d\Omega  
\,\frac{|T(\pi
  N\to \rho N)(s',m^2)|^2}
{s'-s+i\epsilon}
+ ... \ ,
\label{dispint}
\end{equation}
\end{widetext}
where now the ellipses stand for the unitarity cut contributions from other channels as well as left-hand cut contributions. First of
all, there is the three--body cut, which drives the inelasticity of the $T$-matrix. To be concrete, we may write
\be
\rho(m^2)&=&-\frac{N}{\pi}\,{\rm Im}\,\frac{1}{m^2-m_\rho^2+im_\rho\tilde\Gamma},\non
\tilde\Gamma&=&\Gamma\,
\frac{\tilde p(m,m_\pi,m_\pi)^{2\ell+1}}{p_0^{2\ell+1}}
\label{firstspec}
\ee
where $p_0$ is the three-momentum at the nominal resonance mass and $N$ is a normalization factor so that Eq.~(\ref{norm}) is
fulfilled. The factor $(\tilde p/p_0)^{2\ell+1}$ accounts for the centrifugal  barrier and $\tilde p$ is the pion momentum in the
$\rho$ rest frame. Note, $\tilde p=p(m,m_\pi,m_\pi)$ at threshold,  $\sqrt{s}=2m_\pi+M_N$, i. e. the $\rho$ is at rest and the $\rho$
rest frame and overall rest frame coincide. Note also the explicit form of Eq.~(\ref{firstspec}) is only for illustration. The
$m$-dependence of the denominator is more complicated in general (see, e.g., the Appendix), but the only property needed in the
following is the presence of poles in the spectral function.

Indeed, the spectral function $\rho(m^2)$ of Eq.~(\ref{firstspec})
contains a pair of poles located at $m^2=m_0^2$, where $m_0$ denotes the pole position of the $\rho$ meson, located in the
complex plane. We may write $m_0 = m_\rho \pm i\Gamma/2$, where $\Gamma$ denotes the width of the
$\rho$--meson. 

\begin{figure}
\begin{center}
 \includegraphics[width=0.24\textwidth]{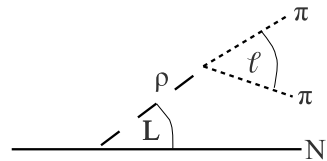}
\caption{The quasi-particle $(\rho)$ coupling to the stable particle $N$ with orbital angular momentum $L$; the decay of the 
quasi-particle into stable particles (2$\pi$) is in $\ell$-wave with respect to the quasi-particle c.m. frame.}
\label{fig:scheme}
\end{center}
\end{figure}

For the existence of branch points in the complex plane, it is sufficient to consider the imaginary part of Eq.~(\ref{dispint}) in the
following, or, more correctly, we consider the analytic function $\delta T$ which is $\delta T={\rm Im}\, T$ for $\sqrt{s}\in
\mathds{R}$, but of course $\delta T\neq {\rm Im}\, T$ for $\sqrt{s}\notin \mathds{R}$ (e.g., $\delta T$ develops an imaginary part for
complex $\sqrt{s}$, whereas ${\rm Im}\, T$ does not). The function $\delta T$ can be straightforwardly evaluated,
\be
\delta\,T&=&-\frac{\pi}{4\sqrt{s}}\int\limits_{4m_\pi^2}^{(\sqrt{s}-M_N)^2} dm^2\,\rho(m^2)\non
&\times&p(\sqrt{s},m,M_N)\,p(\sqrt{s},m,M_N)^{2L}\,g(\sqrt{s},m)\,\,\,
\label{imt}
\ee
with $p$ from Eq.~(\ref{qcm}). In Eq.~(\ref{imt}), we have explicitly denoted a factor of $p^{2L}$ that comes from the $L=0,1,\cdots$
transition $T(\pi  N\to \rho N)$. The function $g(\sqrt{s},m)$ contains the integral $\int d\Omega$ over the part of $|T|^2$ without
these centrifugal barrier factors. In general, $g(m+M_N,m)\neq 0$. The overall process we consider here as an example is shown in
Fig.~\ref{fig:scheme}.

A function $f(\sqrt{s})$ has a branch point $z_b$ at $\sqrt{s}=z_b$, whenever in its integral representation $f(\sqrt{s})=\int_a^b dq\,
\tilde f(\sqrt{s},q)$ the function $\tilde f$ has a simple pole at $q=q_0$ and a $\sqrt{s}=z_b$ exists such that $q_0=a$ or $q_0=b$. For
example, the integrand of the two-body phase space integral $\int_0^\infty dq\, q^2/(\sqrt{s}-E_1-E_2+i\epsilon)$, where
$E_i=\sqrt{m_i^2+q^2}$, has a simple pole at $q_0=p(\sqrt{s},m_1,m_2)$ with the on-shell momentum $p$ from Eq.~(\ref{qcm}). Then, the
branch point is given for the $\sqrt{s}$ for which $q_0=0$ (lower integration limit). This is the case for  $\sqrt{s}\equiv
z_b=m_1+m_2$, i.e. the branch point is at the two-body threshold.

With this knowledge, it is straightforward to determine the branch points of $\delta\,T$: as discussed before, the simple poles of the
integrand (spectral function) are located at the complex $m^2=m_0^2$ which equal the upper integration limit of Eq.~(\ref{imt}) for
$\sqrt{s}=M_N+m_0$. 

Thus, without loss of generality, we have shown that poles in the spectral function at $m=m_0$ lead to branch points of the amplitude at
the complex scattering energy $\sqrt{s}=M_N+m_0$ or 
\be
\sqrt{s}\equiv z_{b1,2}=M_N+m_\rho\pm  i\Gamma/2 \ .
\label{brapopos}
\ee
More general, the model-independent result is that $z_b$ is given by the sum of the mass of the stable particle plus $m_0$, where $m_0$
is the pole position in the scattering amplitude of the subsystem, in this case given  by $\pi\pi$ which resonates through a $\rho$
meson. Eq.~(\ref{brapopos}) has also be obtained in Ref.~\cite{Doring:2009yv}, starting from an explicit expression for the $\pi\pi N$
system, derived from field theory, and in which the $\pi\pi$ subsystem is boosted. In Appendix~\ref{sec:appjm} we will come back to the
connection of that formalism to the present one.

\begin{figure}
\includegraphics[width=0.4\textwidth]{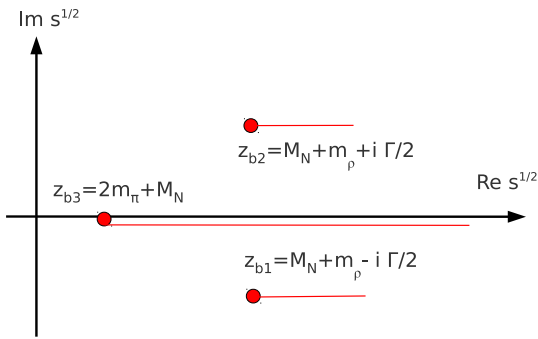}
\caption{Analytic structure of the amplitude. There are three branch points $z_{b1}$, $z_{b2}=z_{b1}^*$, and $z_{b3}$. $z_{b1}$ and 
$z_{b2}$ are structures in $\delta T$ and thus on the second sheet.}
\label{fig:structure}
\end{figure}

The branch points $z_b$ in Eq.~(\ref{brapopos}) have been obtained by considering the upper integration limit in Eq.~(\ref{imt}).
However, also the lower integration limit can coincide with a singularity for a certain $\sqrt{s}$: this is the case for 
\be
\sqrt{s}\equiv
z_{b3}=2m_\pi+m_N
\label{z3}
\ee
 for which the lower integration limit coincides with the branch point singularity coming from the factors of $p$ in the integrand. The
overall analytic structure is shown in Fig.~\ref{fig:structure}.  
The first, physical sheet has the branch point $z_{b3}$ with an associated cut. If the cut is chosen along the real $\sqrt{s}$ axis like
in the figure, the discontinuity of the amplitude is given by $2\,\delta T$ from Eq.~(\ref{imt}). The branch points $z_{b1}$ and
$z_{b2}$ are in $\delta T$, i.e. on the sheet that is obtained by analytically continuing the discontinuity of the first sheet. They
are, thus, on the second sheet, where also resonance poles are normally situated. The branch points $z_{b1}$ and $z_{b2}$ induce the new
sheets 3 and 4; they are analytically connected to the second sheet along the cuts induced by  $z_{b1}$ and $z_{b2}$. In
Fig.~\ref{fig:structure} these cuts are chosen parallel to the real $\sqrt{s}$ axis; in Ref.~\cite{Doring:2009yv} they are chosen
parallel to the imaginary $\sqrt{s}$ axis, which is a convenient choice to search for poles. For the numbering of sheets, see also
Ref.~\cite{Doring:2009yv}.


\subsection{Threshold behavior}
\label{sec:threshold}
Apart from determining the existence and position of branch points, one can also deduce their threshold behavior, i.e. the functional
form of $\delta T$ close to the three $z_b$. In Fig.~\ref{fig:scheme}, the three-body decay is schematically shown. Let the
quasi-particle $(\rho)$ couple to the stable particle $(N)$ in $L$-wave in the overall c.m. system, while the quasi-particle decays into
stable particles (2 pions) in $\ell$-wave with respect to the quasi-particle c.m. frame.

In the following we will use the explicit form of Eq.~(\ref{firstspec}) to determine the threshold behavior. It is clear, however, that
the final results do not depend on this particular form for the spectral function, but only on the fact that the spectral function has
poles [right side of Eq.~(\ref{spectral2})] and the presence of factors of $p$ in Eq.~(\ref{imt}) that follow from the previously given
phase space derivation.
 
To study the behavior of the amplitude in the complex energy plane close to the branch points $z_{b1,2}$, 
complex values of $m^2$ will be needed, and thus the (non-analytic) function ${\rm Im}$
in Eq.~(\ref{firstspec}) needs to be evaluated to obtain a meromorphic expression,
\be
\rho(m^2)=\frac{N}{\pi}\,\frac{m_\rho\tilde \Gamma}{\left(m^2-m_\rho^2\right)^2+m_\rho^2\tilde\Gamma^2}
\,\xrightarrow{m\to m_0}\,
\frac{\Gamma\,h_1(m^2)}{m^2-m_0^2} \ . \nonumber \\
\label{spectral2}
\ee
The right-hand side shows the behavior of $\rho(m^2)$ close to the pole at $m=m_0$; the function $h_1$ does not contain any poles or
zeros close to $m_0$ and thus does not influence the threshold behavior. In particular, $\tilde p^{2\ell+1}$ that appears in the
numerator [cf. Eq.~(\ref{firstspec})], has no zero close to $m_0$ and can be absorbed in $h_1$. Thus the threshold dependence of the
branch points $z_{b1}$ and $z_{b2}$ does not depend on $\ell$, which may appear a surprising result.

\begin{figure}
\begin{center}
 \includegraphics[width=0.47\textwidth]{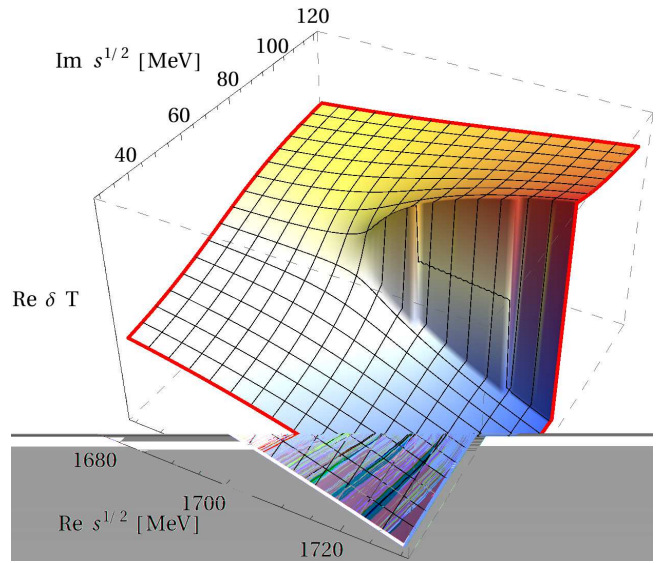}
\caption{Branch point $z_{b2}$ in ${\rm Re}\,\delta T$ in the upper $\sqrt{s}$ half plane, for a realistic $\rho N$ intermediate state.
 The cut is chosen here in the positive ${\rm Re}\,\sqrt{s}$ direction.}
\label{fig:brapo}
\end{center}
\end{figure}

To obtain the threshold behavior of the branch points
$z_{b1,2}$ in the complex plane, one inserts the right-hand side of Eq.~(\ref{spectral2}) into Eq.~(\ref{imt}),
\be
\delta T\sim \int\limits_{4m_\pi^2}^{(\sqrt{s}-M_N)^2} dm^2 
\frac{\Gamma\,(m^2-m_0^2)^{\frac{2L+1}{2}}\,h_2(m^2)}{m^2-m_0^2}
\label{spectral3}
\ee
where we have expanded the argument of the square root of the $p$ factors of Eq.~(\ref{imt}) in $m^2$, at the point
$p(\sqrt{s}=z_b=M_N+m_0,m,M_N)$ to obtain the power of the leading zero from these factors. The function $h_2$ is again analytic, free
of zeros close to $m=m_0$, and does not influence the threshold behavior. The integral may now be evaluated setting this numerator and
$h_2$ constant. The result for the threshold behavior of the branch points $z_{b1,2}$ is [see also Eq.~(\ref{brapopos})]
\be
\delta T(z_{b1,2})\sim \left(\sqrt{s}-z_{b1,2}\right)^{\frac{2L+1}{2}}\sim p(\sqrt{s},m_0,M_N)^{2L+1}  .\non
\ee 
In Fig.~\ref{fig:brapo} we show the branch point $z_{b2}$ in the upper $\sqrt{s}$ half plane [see Fig.~\ref{fig:structure}] for a
realistic $\rho N$ intermediate state and $L=0$. The branch point is clearly visible, together with the cut that in this picture is
chosen in the positive ${\rm Re}\, \sqrt{s}$ direction.

To obtain the threshold behavior for the third branch point at $z_{b3}=2m_\pi+M_N$ [see Fig.~\ref{fig:structure}], we inspect again
Eq.~(\ref{spectral2}). As discussed following Eq.~(\ref{firstspec}), close to $\sqrt{s}=z_{b3}=2m_\pi+M_N$ [see Eq.~(\ref{z3})] the
$\rho$  c.m. frame coincides with the overall c.m. frame, i.e. $\tilde p=p(m,m_\pi,m_\pi)$, and thus the $\ell$-wave decay in the $\rho$
subsystem is also an $\ell$-wave  decay in the overall c.m. system. For $\sqrt{s}$ in the vicinity of $z_{b3}$, the denominator of
Eq.~(\ref{spectral2}) is free of zeros; however, in contrast to the case of $z_{b1,2}$, the numerator $\tilde \Gamma\sim \tilde
p^{2\ell+1}= p^{2\ell+1}=(m^2-4m_\pi^2)^{(2\ell+1)/2}$ does have a zero that contributes to the threshold behavior. Inserting
Eq.~(\ref{spectral2}) (including this factor) in Eq.~(\ref{imt}) and expanding the arguments of the square roots of the $p$ factors
around the zero [cf. Eq.~(\ref{qcm})] one obtains
\be
\delta T&\sim& \int\limits_{4m_\pi^2}^{(\sqrt{s}-M_N)^2} dm^2 \Gamma\left(m^2-4m_\pi^2\right)^{\frac{2\ell+1}{2}}\non
&\times&(m^2-4m_\pi^2)^{\frac{2L+1}{2}}\,h_3(m^2) \ ,
\label{spectral4}
\ee
with a function $h_3$ free of zeros and poles in the vicinity of $z_{b3}$. Integration leads now to the threshold behavior of
$z_{b3}$,
\be
\delta T(z_{b3})&\sim& \left(\sqrt{s}-(2m_\pi+M_N)\right)^{\ell+L+2}\non
&\sim& p(\sqrt{s},M_N,2m_\pi)^{2\ell+2L+4} \ .
\ee 
This corresponds to the opening of the three-body threshold. Note that even if $\ell=L=0$, the threshold behavior is still $\sim p^4$,
i.e. the standard three-body phase space; thus, this threshold opening is always smooth.


\subsection{The limit of vanishing width}
It is instructive to study the limit of a vanishing width of the
$\rho$--meson in Eq.~(\ref{imT}).
Then 
$$
\rho(m^2) \ {{\longrightarrow}} \ \delta (m^2-m_\rho^2) \ \ \mbox{for
}\Gamma\to 0
\ .
$$
This allows us to perform the $m^2$ integration to get
\begin{multline}
\frac1{i}\left(T(\pi N\to \pi N)-T^\dagger(\pi N\to \pi N)\right) \\ 
= (2\pi)^7
\Theta((\sqrt{s}-M_N)^2-m_\rho^2)\int d\Phi_2(P;q,p_3)\\
\times\,  |T(\pi N\to \rho N)(s,m_\rho^2)|^2 
 + ...  \ ,
\label{imTstab}
\end{multline}
such that Eq.~(\ref{dispint}) reduces to the dispersion integral over the
standard two-body cut
\begin{multline}
T(\pi N\to \pi N) \to \frac1{4} \int_{(M_N+m_\rho)^2}^\infty
\frac{ds'}{\sqrt{s'}}\,p(\sqrt{s'},m_\rho,M_N)\\ \times\,
\int d\Omega\frac{|T(\pi
  N\to \rho N)(s',m_\rho^2)|^2}
{s'-s+i\epsilon}
+ ... \ .
\label{dispintstab}
\end{multline}
The imaginary part which is given by 
\be
\delta\, T_{\Gamma\to 0}=-\pi/(4\sqrt{s})\,p(\sqrt{s},m_\rho,M_N)^{2L+1}\,g(\sqrt{s},m_\rho) 
\label{imstable}
\ee
has a branch point at $\sqrt{s}=m_\rho+M_N$, which is simply the ordinary two-body threshold on the real $\sqrt{s}$ axis. As $\Gamma\to
0$, the two branch points $z_{b1,2}$ in the complex plane move towards the real $\sqrt{s}$ axis until they coincide and form this single
branch point at $\sqrt{s}=m_\rho+M_N$. Note that there is a factor of $\Gamma$ in the numerator of Eq.~(\ref{spectral3}), but in the
limit $\Gamma\to 0$, another factor $\sim\Gamma$ appears in the denominators from the two poles moving to the real axis, that cancels
the $\Gamma$ of the numerator. Thus, indeed the branch point persists in the limit $\Gamma\to 0$ with the result given in
Eq.~(\ref{imstable}).

For the third branch point at $z_{b3}=2m_\pi+M_N$, Eq.~(\ref{spectral4}) shows that there are no poles that can prevent the term from
disappearing in the limit $\Gamma\to 0$; thus, as $\Gamma\to 0$, the third branch point fades away. In other words, $\Gamma\to 0$ means
that the $\rho$ decouples from $\pi\pi$ and thus, in our example, the $\pi\pi N$ channel decouples from $\pi N$.


\section{The relevance of branch points in the complex plane}
\label{sec:relevance}
As shown in the previous section, whenever there is a multi-particle intermediate state with pairwise strong correlations, unavoidably
branch points show up in the complex plane. As we will demonstrate on a particular example in this section, their  influence on the data
might well be visible. However, as will be also shown, it is in general not possible to deduce the origin of such a structure from
elastic data only.

The first model we use is the so-called J\"ulich model~\cite{Krehl:1999km,Gasparyan:2003fp,Doring:2009bi,Doring:2009yv,Doring:2010ap}.
It is a coupled channel meson exchange model including the channels $\pi N,\, \eta N,\,K\Lambda,\,K\Sigma$ as well as 3 effective
$\pi\pi N$ channels, namely $\pi\Delta$, $\sigma N$, and $\rho N$. All these two-pion channels show the mentioned kind of branch
points~\cite{Doring:2009yv}. In Appendix~\ref{sec:appjm} we show the connection of the formalism of the J\"ulich model to the
one of the previous section. The J\"ulich model allows for a good description of the available $\pi N$ data in all partial waves
with $j\leq 3/2$ up to an energy of 1.8 GeV and has been recently extended to higher energies, partial waves, and additional
reactions~\cite{Doring:2010ap}.

\begin{figure}
\begin{center}
\includegraphics[width=0.44\textwidth]{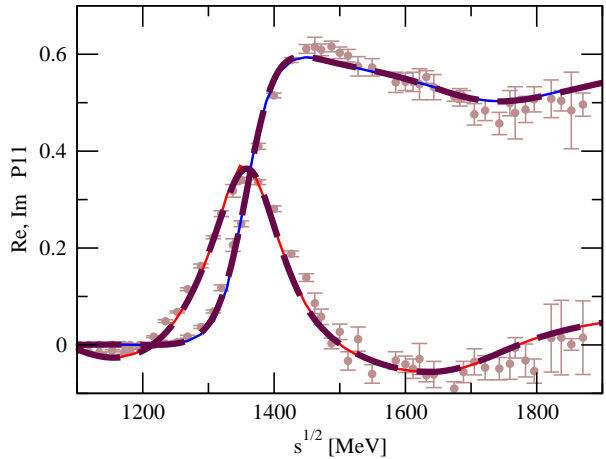}
\caption{Fit of the CMB Zagreb model (solid lines) to the P11 amplitude provided by the J\"ulich model (dashed lines). The ``data'' 
points represent the Single Energy Solution of the GWU/SAID group~\cite{Arndt:2008zz}.}
\label{fig:zagreb}
\end{center}
\end{figure}

To be specific we will focus here on the P11 partial wave and the region around $\sqrt{s}\sim 1.7$ GeV. In this energy region, around
300 MeV above the Roper resonance, signals for another resonance, $N(1710)P_{11}$, have been found in several
analyses~\cite{Nakamura:2010zzi}. It is, however, remarkable that in recent analyses of the GWU/SAID group~\cite{Arndt:2008zz}, there is
no sign for this resonance any more. Like the GWU/SAID analysis, the J\"ulich model contains explicitly the branch points $z_{b1,2}$ in
the complex plane at $\sqrt{s}=M_N+m_\rho\pm i\Gamma/2\sim 1700\pm 75\,i$ MeV. However, there are no poles around these energies (the
only genuine pole term in the P11 partial wave is the nucleon, while the poles of the Roper resonance are dynamically
generated~\cite{Krehl:1999km}). For the purpose of this study we have slightly changed the parameters of the model compared to the
results of Ref.~\cite{Gasparyan:2003fp} to obtain a good description of the GWU/SAID solution. This is shown in Fig. \ref{fig:zagreb} by
the dashed lines. The important point here is that the theoretical amplitude in the complex plane around $\sqrt{s}\sim 1.7$~GeV is free
of poles, but there is the $\rho N$ branch point. 
 
To illustrate the difficulties in determining the origin of structures in the amplitude we fit this J\"ulich model amplitude with
another model, which does not contain the $\rho N$ branch point in the complex plane. For this, we use a Carnegie-Mellon-Berkeley (CMB)
type of model that has been developed by the Zagreb group~\cite{Batinic:1995kr,Batinic:1997gk,Batinic:1998,Batinic:2010}. In this
unitary coupled channel model which respects analyticity, background plus resonances are provided, but all branch points are on the real
axis. The result of the fit, using two resonance terms, is shown in Fig.~\ref{fig:zagreb} by the solid lines.
As the figure shows, the fit is very precise and, in particular, shows no visible discrepancy to the amplitude of the J\"ulich model in
the energy range shown.

However, the behavior in the complex plane is quite different: as mentioned before, there is no complex branch point in the CMB fit by
construction; instead, a pole is found at $1698-130\,i$ MeV which in this case might simulate the branch point missing in that model. 

Thus, at a realistic scale of precision, the $\rho N$ branch point does not manifest itself in a unique structure on the physical axis;
it can be simulated by resonance terms that produce poles in the complex plane. Still, the $\rho N$ branch point is a required structure
of the $S$-matrix, as shown in this study, and we have demonstrated that in an analysis of partial waves, this and other branch points
have to be included to avoid false resonance signals, which of course can totally distort the spectrum of excited baryonic resonances.

In such circumstances, one clearly has to consider other final states in which the resonance candidate shows a clearer signal. As
already proposed in Ref.~\cite{Svarc:2006}, performing global analyses of many different reaction channels within one theoretical ansatz
is a much cleaner way to determine the resonance spectrum than increasing the precision of a partial wave for one reaction.
 
First steps within the coupled-channel J\"ulich model have been undertaken in this direction through the inclusion of some $\rho N$
data~\cite{Krehl:1999km}, $\eta N$ data~\cite{Gasparyan:2003fp}, and, most recently, $K^+\Sigma^+$ data~\cite{Doring:2010ap}. For the
isospin $I=1/2$ sector, we expect the inclusion of $K^0\Lambda$ data to further clarify the role of the $N(1710)P_{11}$ [see also
Ref.~\cite{Ceci:2006ra}].

Thus, the aim of the present short exercise is not to discard the existence of the much-debated $N(1710)P_{11}$ as such; rather, we have
shown that branch points in the complex plane are relevant; in their absence, resonances may be needed to simulate them, and, thus, the
extracted baryon spectrum can be easily distorted. 


\section{Conclusions}

Using only general properties of the $S$-matrix we have shown the existence and determined the position of three branch points induced
by intermediate quasi-two body states. Those are three-body states in which two particles are so strongly correlated that the scattering
amplitude of this subsystem has a pole. A pole in the subsystem necessarily leads to the appearance of branch points in the complex
$\sqrt{s}$ plane of the overall $\pi N$ amplitude. This result is model-independent because it does not depend on any particular
parameterization, but only on analyticity and general properties of the three-body phase space. We have also determined the threshold
behavior of all branch points, which depends on the orbital angular momenta of the two decay processes involved. Finally, on the example
of the $P11$ partial wave, it has been shown that branch points in the complex plane are relevant in partial wave analysis: if a
theoretical amplitude does not contain the branch points, false resonance signals may be obtained. To allow for a reliable extraction of
the baryon spectrum, it is thus mandatory to include also these branch points in the analysis.


\vspace*{0.3cm}
\noindent {\bf Acknowledgements} This work is supported by the DAAD (Deutscher Akademischer Austauschdienst) grant No. D/08/00215. It is
also supported by the DFG (Deutsche For\-schungs\-ge\-mein\-schaft, Gz.: DO 1302/1-2 and SFB/TR-16) and by the EU-Research
Infrastructure Integrating Activity ``Study of Strongly Interacting Matter" (HadronPhysics2, grant n. 227431) under the Seventh
Framework Program of the EU.


\appendix
\section{Spectral representation of the J\"ulich model}
\label{sec:appjm}
In this Appendix the connection of the field theoretical formalism, used in the J\"ulich model of hadron exchange, to the formalism used
in this study is outlined, up to overall normalization factors. For further details of the formalism used in the J\"ulich model, we
refer to Ref.~\cite{Doring:2009yv}. For the example of the $\rho N$ propagator, that is considered here, the propagator on the real axis
is given by
\be
g_{\rho N}(\sqrt{s},k)= \frac{1}{\sqrt{s}-E_N(k)-E_\rho^0(k)-\Sigma(z_\rho(\sqrt{s},k),k)}\non
\label{dyson}
\ee
where $E_N$ is the nucleon energy, $E_\rho^0$ is the $\rho$ energy using the bare $\rho$ mass and $\Sigma$ is the $\rho$ self
energy, where $z_\rho(\sqrt{s},k)$ is the boosted energy for the $\rho$ subsystem. The explicit form of $z_\rho(\sqrt{s},k)$ is quoted
in Ref.~\cite{Doring:2009yv} but for the present discussion the only needed property is that  $z_\rho(\sqrt{s},k=0)=\sqrt{s}-M_N$. The
propagator $g_{\rho N}$ is iterated in the multichannel scattering equation, but to investigate the analytic structure it is sufficient
to consider the one-loop amplitude
\be
G_{\rho N}(\sqrt{s})=\int\limits_0^\infty dk\,k^2\,g_{\rho N}(\sqrt{s},k)
\ee
where for simplicity we have omitted the form factors that regularize this divergent expression. One can rewrite the Dyson-Schwinger
representation of Eq.~(\ref{dyson}) with the spectral function 
\be
S(\omega,k)=-\frac{1}{\pi}\,{\rm Im}\,\frac{1}{\omega-E_N(k)-E_\rho^0(k)-\Sigma(z_\rho(\omega,k),k)}\non
\label{speck}
\ee
resulting in the Lehmann representation 
\be
g_{\rho N}(\sqrt{s},k)=\int\limits_{2m_\pi+M_N}^{\infty}d\omega\,\frac{S(\omega,k)}{\sqrt{s}-\omega+i\epsilon} \ .
\ee
For the imaginary part of the $\rho N$ loop $G_{\rho N}(\sqrt{s})$, one obtains:
\begin{multline}
{\rm Im}\, G_{\rho N}(\sqrt{s})={\rm Im}\int\limits_0^\infty dk\,k^2\,g_{\rho N}(\sqrt{s},k) \\
= -\pi\int\limits_0^\infty dk\, k^2 S(\sqrt{s},k)= -\pi\int\limits_0^{k_1} dk\, k^2 S(\sqrt{s},k)\ .
\label{lalala}
\end{multline}
The last equality shows that the integration can be cut at $k=k_1$ as for $k>k_1$ the spectral function is zero because then
$z_\rho(\sqrt{s},k)<2m_\pi$. In particular, $k_1$ is given by $z_\rho(\sqrt{s},k_1)=2m_\pi$. Note that the explicit
evaluation of the integration limits as done here is necessary if one wants to use the spectral representation in the complex $\sqrt{s}$
plane. This has been shown recently in the context of Feynman parameterized loops~\cite{Doring:2010rd}: the integration limits have to
be analytically continued for complex $\sqrt{s}$ to obtain the analytic continuation of the loop itself, and for this they need to be
known explicitly.

Eq.~(\ref{lalala}) can be rewritten as
\be
{\rm Im}\, G_{\rho N}
(\sqrt{s})&=&\int\limits_{m_1}^{\sqrt{s}-m_\pi}dm\,\frac{S(\sqrt{s},k^{{\rm on}}(m))\,m}{E_m^{{\rm on}}}\non
&\times&(-\pi)\,\frac{k^{{\rm on}}(m)\,
E_\pi^{\rm on}\,E_m^{\rm on}}{\sqrt{s}}
\label{alt}
\ee
with
\begin{multline}
k^{\rm on}(m)=p(\sqrt{s},m,m_\pi),\quad
E_\pi^{\rm{on}}=\sqrt{m_\pi^2+(k^{\rm on})^2},\\
E_m^{\rm{on}}=\sqrt{m^2+(k^{\rm on})^2}.
\label{kon}
\end{multline}
and $p$ from Eq.~(\ref{qcm}). The lower integration limit $m_1$ is given as the solution of $z_\rho(\sqrt{s},k^{\rm on}(m_1))=2m_\pi$.
The second fraction in Eq. (\ref{alt}) can be compared to the imaginary part of the well-known~\cite{Doring:2009yv} propagator of two
stable particles $M$ and $N$,
\be
{\rm Im}\,G_{\rm stable}=-\pi\, \frac{k^{{\rm on}}(m=\sqrt{s})\,E_M^{\rm on}\,E_N^{\rm on}}{\sqrt{s}} \ .
\label{stable}
\ee
Thus, the imaginary part of a loop with one stable and one unstable particle can be expressed as an integral over a distribution of
imaginary parts of the form of Eq.~(\ref{stable}). Comparing Eq.~(\ref{alt}) to Eq.~(\ref{imt}), one sees the formal similarity: there
is an integral of a spectral function, that has poles [cf. Eq.~(\ref{speck})], together with the factor $k^{{\rm on}}(m)
=p(\sqrt{s},m,m_\pi)$, and both ingredients produce the three branch points $z_{b1,2,3}$ as has been shown in the main text (we have
omitted here the additional $2L$ powers of $p$ for simplicity). There is a difference in the chosen parameterization in terms of the
spectral function [compare $p(\sqrt{s},m,m_\pi)$ in Eq.~(\ref{kon}) vs. $p(\sqrt{s},m,M_N)$ in Eq.~(\ref{imt})], but this does not
change the position of the branch points. 

Indeed, $k^{\rm on}=0$ for the upper integration limit $m=\sqrt{s}-m_\pi$ and thus $z_\rho(\sqrt{s},0)=\sqrt{s}-M_N$. The poles of the
$\rho$ resonance in the spectral function $S$ are located at the complex $z_\rho=z_\rho^0$ and consequently the integration limit equals
the pole position for $\sqrt{s}\equiv z_{b1,2}=M_N+z_\rho^0$ which is indeed Eq.~(\ref{brapopos}).  The singularity at $k^{{\rm on}}=0$,
coming from the factor $k^{{\rm on}}(m)$ in Eq.~(\ref{alt}), is also reached if $m=\sqrt{s}-m_\pi$. It is easy to show that this $m$
equals the lower integration limit $m_1$ for $\sqrt{s}=2m_\pi+M_N$ and thus Eq.~(\ref{alt}) indeed provides also the third branch point
$z_{b3}$ from Eq.~(\ref{z3}).

In Ref.~\cite{Doring:2009yv}, the amplitude has been analytically continued to the complex plane using contour deformation. In fact, one
could use the representation of Eq.~(\ref{alt}) for the same purpose in principle. As shown in the main text, for this, one has to
respect the analyticity of the spectral function, i.e. the ${\rm Im}\,()$ function in Eq.~(\ref{speck}) needs to be explicitly evaluated
like in Eq.~(\ref{spectral2}). Second, and this is an additional complication, the self energy $\Sigma$ itself has a two-sheet structure
and the corresponding cut needs to be rotated as specified in Ref.~\cite{Doring:2009yv}. This cut in $\Sigma$ induces the cut of branch
point $z_{b3}$ in the overall $\pi N$ amplitude. Apart from this and a carefully chosen integration path for the $m$-integration of
Eq.~(\ref{alt}), there are no additional complications, and the spectral representation allows for an alternative way of analytic
continuation.


\end{document}